\documentclass[12pt,a4paper]{article}
\usepackage{enumerate} 
\usepackage{euscript}
\usepackage{amsmath,amssymb,amsthm}
\usepackage{fullpage,hyperref}

\def\ca#1{{\cal#1}}
\def\scri#1{{\EuScript#1}}
\newcommand\OO{{\cal O}}
\newcommand\FF{\mathbb{F}}
\newcommand\sem{\setminus}
\def\mx#1{\mbox{\boldmath$#1$}}
\def\of#1{{\EuScript#1}}
\newcommand\pw{\operatorname{pw}}
\newcommand\bw{\operatorname{bw}}

\newtheorem{theorem}{Theorem}[section]
\newtheorem{lemma}[theorem]{Lemma}
\newtheorem{corollary}[theorem]{Corollary}
\newtheorem{alg}[theorem]{Algorithm}
\newtheorem{claim}[theorem]{Claim}

\newcommand{\TheTitle}{A Simpler Self-reduction Algorithm for \mbox{Matroid Path-width}} 

\title{{\bf\TheTitle}\thanks{
	{Supported by the project 17-00837S
        	of the Czech Science Foundation.}}}

\author{
  Petr Hlin\v{e}n\'y\\\normalsize
	Faculty of Informatics, Masaryk University,
				Brno, Czech Republic
    	\\\mbox{\small\texttt{hlineny@fi.muni.cz}}
}

\begin{document}

\maketitle

\begin{abstract}
Path-width of matroids naturally generalizes the better known parameter of 
path-width for graphs, and is NP-hard by a reduction from the graph case.
While the term {\em matroid path-width} was formally introduced by
Geelen--Gerards--Whittle [JCTB 2006] in pure matroid theory, 
it was soon recognized by Kashyap [SIDMA 2008]
that it is the same concept as long-studied so called trellis complexity
in coding theory, later named {\em trellis-width},
and hence it is an interesting notion also from the algorithmic perspective.
It follows from a result of Hlin\v{e}n\'y [JCTB 2006] that the decision
problem, whether a given matroid over a finite field has path-width at most
$t$, is fixed-parameter tractable (FPT) in~$t$, but this result does not give any clue
about constructing a path-decomposition.
The first constructive and rather complicated 
FPT algorithm for path-width of matroids over a finite
field was given by Jeong--Kim--Oum [SODA 2016].
Here we propose a simpler ``{self-reduction}'' FPT algorithm for
a path-decomposition.
Precisely, we design an efficient routine that
constructs an optimal path-decomposition of a matroid
by calling any subroutine for testing whether the path-width of a matroid is at most~$t$
(such as the aforementioned decision algorithm for matroid path-width).

\end{abstract}


\section{Introduction}

An ordinary path-decomposition of a graph $G$,
see \cite{rs83-GMI}, is a sequence of sets 
$\big(X_i\subseteq V(G): i=1,\dots,p\big)$, such that;
(i) $\bigcup_{i=1}^p X_i=V(G)$ and for every $1\leq i<j<k\leq p$,
we have $X_j\supseteq X_i\cap X_k$, and
(ii) for every $e=uv\in E(G)$ there is $1\leq i\leq p$ 
such that~$u,v\in X_i$.
The width of this decomposition equals $\max_{1\leq i\leq p}|X_i|-1$,
and the {\em path-width} of $G$ is the minimum width over all 
path-decompositions of~$G$.
This notion, together with related tree-width, has received great attention
in the Graph Minors project of Robertson and Seymour.

There is another, more recent view of path-width;
the {\em matroid path-width} defined first by Geelen, Gerards and
Whittle~\cite{ggw06} in matroid research.
We refer to Section~\ref{sec:prelim} for the definition.
While the two variants of path-width are indeed tightly related,
there is no simple explicit formula between the ordinary
path-width and the matroid path-width of the same graph.
Matroid path-width of graphs has been recently studied in some papers,
e.g.~\cite{kty14}.
Our interest in matroid path-width, however, lies beyond the graph case.

A similar notion to path-width has been considered for quite some time also
in the area of coding theory, under various names such as the ``trellis
complexity'' of a code, e.g.~\cite{vardy98,jmv98}.
In 2008, Kashyap~\cite{kashyap08} observed that this is the same parameter
as the aforementioned path-width~\cite{ggw06} of a vector matroid 
represented by the generator matrix of a linear code.
He introduced for it the new name {\em trellis-width} of a linear code,
and proved that computing trellis-width is NP-hard by a reduction from
graph path-width.
Kashyap also asked, as one of the main open problems in~\cite{kashyap08},
how difficult it is to decide whether the trellis-width of a linear code
over a fixed finite field is at
most~$t$, and to construct the corresponding optimal decomposition in the
{\sc Yes} case, where $t\in \mathbb N$ is a fixed parameter.

Concerning the first half of Kashyap's question, the decision problem is in FPT 
({\em fixed-parameter tractable}) which follows already from the author's
papers~\cite{hlineny05,hli06}.
Recall that a parameterized problem is in FPT if it admits an algorithm 
with runtime of order $\OO(f(t)\cdot n^c)$ where $t$ is the parameter, $n$
the input size and $c$ a constant.
We briefly sketch two key ideas on which an FPT algorithm for
deciding `trellis-width $\leq t$' is based
(see Section~\ref{sec:improve} for full details):
\begin{itemize}
\item The branch-width of the underlying vector matroid of a linear code is
upper-bounded in terms of $t$, the assumed trellis-width bound.
Hence there are only finitely many ``minimal obstructions''
for the property `trellis-width $\leq t$' for each $t\in \mathbb N$
and each finite field $\FF$,
which follows from~\cite{ggw02}.
(A similar observation occurs also in Kashyap~\cite{kashyap08}.)
\item For bounded branch-width of a vector matroid, over any finite field~$\FF$,
we can construct an approximate branch-decomposition of it in FPT,
see~\cite{hlineny05}.
Then we can, again in FPT, check presence of each one of these finitely many
obstructions, see~\cite{hlin03-mfcs,hli06}.
\end{itemize}
A careful reader may immediately notice a problem of the suggested scheme---in
what way can we get a corresponding trellis- or path-decomposition from it?
The sad truth is that in {\em no way}.
To get a corresponding decomposition, a new approach is needed.

Speaking in general, situations in which we get an algorithm which efficiently
computes the value of a solution to a certain problem, but not the
witnessing solution, are not common in algorithm theory, however, they
are also not rare.
In such situations, the so-called {\em self-reduction} routine helps,
that is, repeated calls to the algorithms for a solution value
(on various inputs derived from the given one)
are used to find an admissible solution of the given instance.
For a brief example, imagine having an oracle for testing $3$-colourability
of any graph; how could we then find an actual $3$-colouring of
a particular graph~$G$ using it?
The corresponding self-reduction algorithm is quite simple:
trying to add new edges to $G$ as long as the oracle certifies that
a $3$-colouring still exists, the final outcome will be a complete
tri-partite graph exhibiting three valid colour classes of~$G$.

\medskip
Our situation is analogous to that of constructing an optimal matroid
branch-decomposition, for which the aforementioned paper~\cite{hlineny05}
provided an approximate construction and an exact decision (the value) in FPT.
Building upon that, Oum and the author~\cite{ho08} later designed a
self-reduction routine which constructs an optimal branch-decomposition
of a matroid over a finite field, by calling the decision subroutine for
exact branch-width.
It appears very natural to try to extend the self-reduction approach of~\cite{ho08} 
also for path-decompositions but this, unfortunately, does not easily work.
Instead, Jeong, Kim and Oum~\cite{jko16,jko16j} designed a rather complicated
standalone algorithm for the construction of an optimal path-decomposition
of a matroid over a finite field, which runs in FPT time for the parameter path-width.
In their algorithm, they refer back to the ideas and techniques of Bodlaender
and Kloks~\cite{bk96} from graphs.

In this paper we complete the whole picture by providing a new
{\em self-reduction routine} for constructing an optimal matroid
path-decomposition, partially inspired by~\cite{ho08}.
That is, our routine uses recursive calls to (any) decision subroutine for
exact path-width to efficiently construct the output path-decomposition.
As the decision subroutine we may use, e.g., the above mentioned FPT algorithm
for matroid path-width over a finite field based on~\cite{hlineny05,hli06}.

In a nutshell, we contribute the following:
\begin{enumerate}
\item
A nonuniform FPT algorithm that, for fixed parameters $t$ and $|\FF|$,
inputs an $n$-element matroid $M$ represented by a matrix over a finite
field~$\FF$, and in $\OO(n^3)$ time
constructs a path-decomposition of $M$ of width $\leq\! t$ or concludes that
the path-width of $M$ is~$>\!t$.
This is {\em not} a better or faster algorithm than
in aforementioned Jeong, Kim and Oum~\cite{jko16,jko16j}, 
but the advantage of our approach is
in much simpler design and proof of the algorithm.
{\em(Section~\ref{sec:selfred} -- Theorem~\ref{thm:main} for a generic
algorithm, and Theorem~\ref{thm:alg-n3} for improved runtime)}
\item
An FPT algorithm that, for a fixed parameter $t$,
a given {\em oracle function $\cal P$} testing if the path-width 
of a matroid is~$\leq\! t$,
and an input $n$-element abstract matroid $M$, 
constructs a path-decomposition of $M$ of width $\leq\! t$ or concludes that
the path-width of $M$ is~$>\!t$.
This part is not achieved by~\cite{jko16,jko16j}.
{\em(Section~\ref{sec:improve} -- Theorem~\ref{thm:alg-abstract})}
\end{enumerate}

Regarding (1.), a `{\em nonuniform FPT algorithm}'
means that there is a sequence of algorithms for each
values of the parameters~$t,|\FF|$, rather than one universal algorithm.
This weakness is only due to the used decision subroutine for matroid path-width
in which we do not know explicitly the finite list of obstructions.
If, on the other hand, a different decision algorithm for matroid path-width
is found in the future (which may be easier than such a constructive
algorithm), or some explicit bound on the path-width obstructions is proved
(as in the case of branch-width~\cite{ggrw03}),
then our results immediately give corresponding uniform FPT algorithms.

\section{Preliminaries}\label{sec:prelim}

We refer to the textbook of Oxley~\cite{Oxley:2006}
for standard matroid material and terminology.

\paragraph{Matroids; rank and connectivity}
A \emph{matroid} is a pair $M=(E,\of B)$ where $E=E(M)$ is the ground
set of $M$ (elements of $ M$),
and $\of B\subseteq2^{E}$ is a nonempty collection of \emph{bases} of $ M$,
no two of which are in an inclusion.
Moreover, matroid bases satisfy the ``exchange axiom'':
if $B_1,B_2\in\of B$ and $x\in B_1\sem B_2$,
then there is $y\in B_2\sem B_1$ such that $(B_1\sem\{x\})\cup\{y\}\in\of B$.
We consider only finite matroids.

All matroid bases have the same cardinality called the \emph{rank} $r(M)$
of the matroid.
Subsets of bases are called \emph{independent},
and sets that are not independent are \emph{dependent}.
Minimal dependent sets are called \emph{circuits}.
The \emph{rank function} $r_{M}(X)$ of $M$ maps subsets of $E(M)$ to
non-negative integers; $r_{M}(X)$ equals the maximum
cardinality of an independent subset of a set $X\subseteq E(M)$.
The rank function is {\em submodular},
meaning that $r_{M}(X)+r_{M}(Y)\geq r_{M}(X\cup Y)+r_{M}(X\cap Y)$
for any~$X,Y\subseteq E(M)$,
and it fully defines a matroid on its ground set.
A matroid $M$ is \emph{uniform} if all subsets of $E(M)$ 
of size equal to $r(M)$ are bases,
and it is also denoted by $U_{r,n}$ where $r=r(M)$ and~$n=|E(M)|$.

For $X\subseteq E$, {\em deletion of $X$} results in the matroid $M\sem X$
which is defined by the restriction of the rank function $r_M$ to~$E\sem X$.
On the other hand, {\em contraction of $X$} results in the matroid $M/X$
which is defined by the rank function 
$r_{M/X}(Y):=r_M(X\cup Y)-r_M(X)$ for all $Y\subseteq E\sem X$.
Matroids of the form $M/X\setminus Y$ are called \emph{minors} of $M$.

\medskip

The {\em closure of a set $X\subseteq E$ in~$M$}, denoted by $cl_M(X)$,
is defined by 
$$
cl_M(X):=\big\{e\in E: r_M(X\cup\{e\})=r_M(X)\big\}
.$$
The closure of $X$, hence, includes all elements dependent on
(or spanned by)~$X$.
Sets $X$ such that $X=cl_M(X)$ are closed, or {\em flats}.

We, moreover, define the (symmetric and submodular) {\em connectivity
function} of $M$ by
$$
	\lambda_{M}(X):=r_{M}(X)+r_{M}(E\sem X)-r(M)
$$
for all subsets $X\subseteq E$.
Any bipartition $(X,Y)$ of $E$ (where $Y=E\sem X$) is called
a {\em separation in $M$} of connectivity value
$\lambda_M(X)=\lambda_M(Y)$, or shortly a {\em$k$-separation}, 
if $\lambda_M(X)=k-1$ and both $|X|,|Y|\geq k$.
Informally, $\lambda_M$ measures how much the two sides of a separation
``share together'' in terms of rank.
A matroid is {\em connected} if and only if it has no $1$-separation.
It is well-known that in a connected matroid, every two elements belong to a
common circuit (this is analogous to graph $2$-connectivity).

We also define the following extension of the connectivity function which
will be useful in our context
$$
\mu_M(X,A):= r_{M}(X\cup A)+r_{M}\big((E\sem X)\cup A\big)-r(M)
.$$
For example, $\mu_M(X,A)=\lambda_M(X)$ if and only if
$A\subseteq cl_M(X)\cap cl_M(E\sem X)$ or, in other words,
if $A$ is spanned by both $X$ and~$E\sem X$.
If $e\in cl_M(X)\cap cl_M(E\sem X)$, then we say that
$e$ is {\em in the guts of the bipartition $(X,E\sem X)$}.
As another example we mention that, if $\mu_M(X,A)=r_M(A)$ then
every element in the guts of $(X,E\sem X)$ belongs to the closure of~$A$.

\paragraph{Matroid path-width \cite{ggw06}}
Let $M$ be an $n$-element matroid.
Any permutation $Y=(e_1,e_2,\dots,e_n)$ of the elements $E(M)$ is called a
{\em path-decomposition} of $M$.
The {\em width of $(e_1,e_2,\dots,e_n)$} is defined
$$
w_M(Y)=w_M(e_1,e_2,\dots,e_n):= \max_{i=1,\dots,n} \lambda_M(\{e_1,\dots,e_i\})
,$$
and the {\em path-width $\pw(M)$ of $M$} is the least width over all
path-decompositions of~$M$, i.e.
$$
\pw(M):= \min_{\text{permut.~}\pi\in S_n} 
	w_M\left( e_{\pi(1)}, e_{\pi(2)},\dots, e_{\pi(n)} \right)
.$$
We say, for any $1\leq i<n$, that the bipartition
$\big(\{e_1,\dots,e_i\},\{e_{i+1},\dots,e_n\}\big)$
is {\em displayed} by the path-decomposition $Y$,
and we refer to $\big(\{e_1,\dots,e_i\},\{e_{i+1},\dots,e_n\}\big)$ as to
the {\em bipartition at position~$i$}.

The notion of matroid path-width is related to the better known
parameter of branch-width.
A tree $T$ is {\em cubic} if its vertex degrees are $3$ or $1$.
A {\em branch-decomposition} of a matroid $M$ is a pair $(T,\tau)$
where $T$ is a cubic tree and $\tau:E(M)\to\ell(T)$ is a bijection of the
elements of $M$ to the leaves of $T$.
Every edge $e\in E(T)$ partitions the leaves of $T$ into two sets $L_1,L_2$,
and we say the bipartition $\big(\tau^{-1}(L_1),\tau^{-1}(L_2)\big)$ 
is {\em displayed by~$(T,\tau)$}.
We define the width of $e$ as $\lambda_M\big(\tau^{-1}(L_1)\big)+1$ and the
width of~$(T,\tau)$ as the maximum of widths over all edges of $T$.
The {\em branch-width} $\bw(M)$ of $M$ is the minimum width over all
branch-decompositions of~$M$.

A cubic tree is a caterpillar if it is obtained by connecting leaves to a path.
{\em Linear branch-width} of a matroid $M$ is defined as ordinary
branch-width with a restriction that the cubic tree $T$ must be a caterpillar.
One can easily observe that this notion coincides with that of matroid
path-width (except the artificial `$+1$' term above);
the path-width of $M$ is always one less than its linear branch-width.
Consequently, we have:

\begin{lemma}\label{lem:pwbw}
For any matroid $M$, we have $\bw(M)\leq\pw(M)+1$.
\qed\end{lemma}

\paragraph{Assorted matroid claims}
We list some elementary and intuitive technical claims about matroids 
which will be used in the proof of our algorithm.

\begin{lemma} \label{lem:circexch-exact}
Let $M$ be a matroid and $C_1,C_2\subseteq E(M)$
be two circuits of $M$ such that 
$|C_1\cap C_2|=1$
and $r_M(C_1)+r_M(C_2)=r_M(C_1\cup C_2)+1$.
Then $C_1\Delta C_2$ (the symmetric difference) is also a circuit of~$M$.
\end{lemma}
\begin{proof}
Let $C_1\cap C_2=\{f\}$.
By the standard circuit exchange axiom there exists a circuit
of $M$ contained in the set $C_3:=(C_1\cup C_2)\sem\{f\}=C_1\Delta C_2$.
Consider any $e\in C_3$ where, up to symmetry, $e\in C_2\sem C_1$.
We have $|C_3\sem\{e\}|=|C_1|-1+|C_2|-1-1=r_M(C_1)+r_M(C_2)-1=r_M(C_1\cup C_2)$.
At the same time, since $C_1,C_2$ are circuits and $f\in C_1\cap C_2$, we
have $r_M(C_1\cup C_2)=r_M\big(C_1\cup(C_2\sem\{e\})\big)=
 r_M\big((C_1\sem\{f\})\cup(C_2\sem\{e,f\})\big)=r_M(C_3\sem\{e\})$
and so $C_3\sem\{e\}$ is independent.
Therefore, $C_3$ itself is the circuit.
\end{proof}

\begin{lemma}\label{lem:onesidesep}
Let $M$ be a matroid and $X\subseteq E=E(M)$.
If $e,f\in E$ such that 
$\mu_M(X,\{e\})=\mu_M(X,\{f\})=\mu_M(X,\{e,f\})=\lambda_{M}(X)+1$,
then either $e,f\in X$ or $e,f\not\in X$.
\end{lemma}
\begin{proof}
Let $Y=E\sem X$.
Assume the contrary, i.e.\ up to symmetry, $e\in X$ and $f\in Y$.
From $r_M(X)+r_M(Y)-r(M)+1 =
	\lambda_{M}(X)+1=\mu_M(X,\{f\})=
	r_{M}(X\cup\{f\})+r_{M}(Y)-r(M)$
we immediately get $r_M(X\cup\{f\})=r_M(X)+1$
and, by symmetry, $r_M(Y\cup\{e\})=r_M(Y)+1$.
This leads to
\begin{eqnarray*}
\lambda_{M}(X)+1 &=& \mu_M(X,\{e,f\})
	=r_{M}(X\cup\{f\})+r_{M}(Y\cup\{e\})-r(M)
\\
	&=& r_M(X)+r_M(Y)+2-r(M) = \lambda_{M}(X)+2
,\end{eqnarray*}
a contradiction.
\end{proof}

\begin{lemma}\label{lem:pwminor}
Let $M$ be a matroid and $N$ a minor of $M$.
Then $\pw(N)\leq\pw(M)$.
\end{lemma}
\begin{proof}
Consider $X\subseteq E(M)$ and $e\in E(M)$.
It is well-known that $\lambda_{M\sem e}(X\sem\{e\})\leq\lambda_M(X)$
and $\lambda_{M/e}(X\sem\{e\})\leq\lambda_M(X)$.
Hence, by induction on $|E(M)|-|E(N)|$, 
the restriction of any path-decomposition of $M$ is a
path-decomposition of~$N$ of at most the same width.
\end{proof}

\begin{lemma}\label{lem:circuitspan}
Let $M$ be an $n$-element matroid and $(e_1,\dots,e_n)$ be a
path-decomposition of $M$ of width $t=w_M(e_1,\dots,e_n)$.
For an index $i$ let $X=\{e_1,\dots,e_i\}$ and
$Y=\{e_{i+1},\dots,e_n\}=E(M)\sem X$ such that $\lambda_{M}(X)=t$.
Assume that there exists a circuit $C\subseteq E(M)$ such that no element of
$C$ is in the guts of $(X,Y)$ and $\mu_M(X,C)=r_M(C)$.
Then $X\cap C\not=\emptyset\not=Y\cap C$.
\end{lemma}

\begin{proof}
We proceed by means of contradiction, aiming to show
that $w_M(e_1,\dots,e_n)>t$.
Up to symmetry, let $Y\cap C=\emptyset$, meaning that~$C\subseteq X$.
Let $j\leq i$ by the largest index such that $e_j\in C$,
and $X'=\{e_1,\dots,e_{j-1}\}$, $Y'=\{e_j,\dots,e_n\}$.
From the assumptions $\lambda_{M}(X)=r_{M}(X)+r_{M}(Y)-r(M)=t$
and $\mu_M(X,C)= r_{M}(X)+r_{M}(Y\cup C)-r(M) =r_M(C)$,
we derive
\begin{equation}\label{eq:CYt}
r_M(C)+r_{M}(Y)-r_{M}(Y\cup C)=t
.\end{equation}
We have $r_M(C\sem\{e_j\})=r_M(C)$ since $C$ is a circuit,
and $r_{M}(Y\cup\{e_j\})=r_{M}(Y)+1$ since $e_j$ is not in the guts of $(X,Y)$.
Hence we can rewrite \eqref{eq:CYt} as
\begin{equation}\label{eq:CYt1}
r_M(C\sem\{e_j\})+r_{M}(Y\cup\{e_j\})-r_{M}(Y\cup C)=t+1
.\end{equation}
Note that $Y\cup\{e_j\}\subseteq Y'$ and $C\sem\{e_j\}\subseteq X'$.
We conclude the proof by showing
\begin{eqnarray*}
\lambda_M(X') &=& r_{M}(X')+r_{M}(Y')-r(M)
	\\ &=& r_{M}(X')+\big( r_{M}(Y')-r_{M}(X'\cup Y') \big)
	\\ &\geq& r_{M}(X')+\big( r_{M}(Y\cup\{e_j\})-r_{M}(X'\cup Y\cup\{e_j\}) \big)
	\\ &=& r_{M}(Y\cup\{e_j\})+\big( r_{M}(X')-r_{M}(X'\cup Y\cup\{e_j\}) \big)
	\\ &\geq& r_{M}(Y\cup\{e_j\})+\big(
			r_{M}(C\sem\{e_j\})-r_{M}((C\sem\{e_j\})\cup Y\cup\{e_j\}) \big)
	\\ &=& r_M(C\sem\{e_j\})+r_{M}(Y\cup\{e_j\})-r_{M}(C\cup Y)=t+1
,\end{eqnarray*}
using submodularity and \eqref{eq:CYt1}.
This however contradicts $w_M(e_1,\dots,e_n)=t$.
\end{proof}

\begin{lemma}\label{lem:circguts}
Let $M$ be a matroid and $Y\subseteq E(M)$, $Y'=E(M)\sem Y$.
Assume that $Q\subseteq E(M)$ is such that all elements of
$Q$ are in the guts of $(Y,Y')$ and $r_M(Q)=\lambda_M(Y)$.
If $C$ is a circuit of $M$ and $e\in C\sem Y$,
then there exists a circuit $C'$ of $M$ 
such that $e\in C'\subseteq (C\sem Y)\cup Q$ and $C'\subseteq cl_M(Y')$.
\end{lemma}
\begin{proof}
If $C\subseteq cl_M(Y')$, we are done.
Otherwise, both the sets $C\cap Y,C\cap Y'$ are nonempty and independent,
and $r_M(Q)>0$.
Let $Q_1\subseteq Q$ be a basis of~$Q$, and
$M'$ be the restriction of $M$ onto $Q_1\cup C$.
We aim to show that the set $D:=(C\sem Y)\cup Q_1$ is dependent.
On the contrary, assume that $D$ is independent and choose
$D_1\subseteq C\sem D$ such that $D\cup D_1$ is a basis of~$M'$.
Let $M_1:=M'/D_1$ be obtained by contracting $D_1$ (hence $M_1$ has a basis~$D$).
Then $C\sem D_1$ is a circuit of $M_1$ and
$\emptyset\not=(C\sem D_1)\cap Y\subseteq cl_{M_1}(Q_1)$.
Consequently,
$\big[(C\sem D_1)\sem\big((C\sem D_1)\cap Y\big)\big]\cup Q_1=
 (C\sem Y)\cup Q_1 =D$ is dependent in $M_1$,
and so it is in~$M$ since $r_{M_1}(D)=r_M(D)$.

Hence, dependent $D$ contains a circuit~$C'$ of~$M$, and
since $Q_1$ is independent, we may choose $C'$ such that~$e\in C'$.
Finally, $C'\subseteq (C\sem Y)\cup Q$ and 
$C'\subseteq cl_M(Y')$ are true by the definition of~$D$.
\end{proof}

\paragraph{Matroid representation and extensions}

A standard example of a matroid is given by a set of vectors
(forming the columns of a matrix $\mx A$) with usual linear independence.
The matrix $\mx A$ is then called a \emph{(vector) representation} of the matroid.
We will consider only representations $\mx A$ over finite fields.
Since non-zero scaling of vectors does not change linear dependencies,
vector representations can also be seen as point configurations in the
projective space over~$\FF$, which will be the view followed throughout this paper.
(Note that parallel vectors are represented by the same points.)

We now briefly illustrate the ``geometric'' meaning of matroid terms.
\begin{itemize}
\item The matroid closure of a set $X$ corresponds to the affine closure or
{\em span $\langle X\rangle$} of the points representing~$X$
(note that considering the points of $X$ in a projective space, 
$\langle X\rangle$ does {\em not} contain the origin {\boldmath$0$}).
The rank of $X$ is the dimension or rank of the span of $X$.
\item For a bipartition $(X,Y)$ of $M$,
the {\em guts} of $(X,Y)$ consists exactly of the points in the intersection 
of the spans of $X$ and~$Y$, that is $\langle X\rangle\cap\langle Y\rangle$,
and $\lambda_M(X)$ is the rank of this guts.
The value of $\mu_M(X,A)$ equals the rank of the space spanned by
$(\langle X\rangle\cap\langle Y\rangle)\cup A$.
\item All the previous entities can be straightforwardly computed 
by means of standard linear algebra over the matrix~$\mx A$.
\end{itemize}

There is one particular operation we need to discuss in close detail.
For a matroid $M$ we say that a matroid $M_1$ is a {\em free extension
of~$M$ by element~$e$} if $e\in E(M_1)$ and $M=M_1\sem e$, $r(M_1)=r(M)$,
and for every $X\subseteq E(M)$ we have $r_{M_1}(X\cup\{e\})=r_M(X)+1$
unless~$r_M(X)=r(M)$.
This is equivalent to claiming that every circuit of $M_1$ containing $e$
has full rank~$r(M_1)$.
Informally saying, $e$ is added to $M$ without any unforced dependency
-- geometrically, in a general position.
We will also say that $e$ is {\em freely placed} in~$M$
(see also (M\ref{it:aclosure}) in Section~\ref{sec:improve}).
We will use the following:

\begin{lemma}\label{lem:extalpha}
Let $M$ be a matroid of rank $r$ represented by a matrix $\mx A$ over a
finite field~$\FF$.
Let $\alpha$ be a root of an irreducible polynomial of degree~$r$
in~$\FF$, and denote by $\mx b=(1,\alpha,\dots,\alpha^{r-1})^T$.
Let $\FF(\alpha)$ be the extension field of $\FF$ obtained by
adjoining~$\alpha$ to~$\FF$.
Then the matrix $[\mx A|\,\mx b]$ over $\FF(\alpha)$ represents
a free extension of $M$ by an element~$b$.
\end{lemma}
\begin{proof}
Assume the contrary, that $\mx b$ is a linear combination
over $\FF(\alpha)$ of the columns of
a column-submatrix $\mx A'\subseteq\mx A$ of rank less than~$r$.
Since $\mx A'$ has $r$ rows denoted by $\mx a'_1,\mx a'_2,\dots,\mx a'_r$,
they are linearly dependent as vectors, and so for some
$\lambda_1,\dots,\lambda_r\in\FF$ (not all~$0$) it holds
$\lambda_1\mx a'_1+\lambda_2\mx a'_2+\dots,\lambda_r\mx a'_r=\mx0$.
However, since $\mx b$ is a linear combination of the columns of $\mx A'$,
we have also 
$\lambda_1+\lambda_2\alpha^1+\dots,\lambda_r\alpha^{r-1}=0$.
This contradicts the assumption that $\alpha$ is a root of an 
irreducible polynomial of degree~$r$ over~$\FF$.
\end{proof}

The next two lemmas cover some simple properties of path-decompositions of
represented matroids.
\begin{lemma}\label{lem:alwaysSigma}
Let $\FF$ be a finite field, $|\FF|\geq3$, and $t\geq2$ be an integer.
Denote by $P$ the point set of some rank-$t$ projective space
$\Sigma$ over~$\FF$.
Then, for any permutation $(p_1,\dots,p_k)$ of $P$ there exists $i$ such
that $\langle p_1,\dots,p_i\rangle=\Sigma=\langle p_{i+1},\dots,p_k\rangle$.
In other words, $w_M(p_1,\dots,p_k)=t$ where $M$ is the matroid represented by~$P$.
\end{lemma}
\begin{proof}
We have $|P|=k=\frac{q^t-1}{q-1}$ points where $q=|\FF|$ ($\!$\cite{Oxley:2006}),
and every proper subspace of $\Sigma$ has at most $k'=\frac{q^{t-1}-1}{q-1}$ points.
Since, by simple calculus, $k'<\lfloor k/2\rfloor$ when $q\geq3$,
we are done by choosing $i=\lfloor k/2\rfloor$.
\end{proof}

\begin{lemma}\label{lem:repl-pw}
For $i=1,2$, let $M_i$ be a matroid represented over a finite field~$\FF$,
and $Y_i$ be a path decomposition of $M_i$ of width at most~$t$.
Assume there exist prefixes $Z_i$ of $Y_i$, $i=1,2$, such that
$\langle Z_1\rangle\cap \langle E(M_1)\sem Z_1\rangle=
 \langle E(M_1)\rangle\cap\langle E(M_2)\rangle
 \subseteq \langle E(M_2)\sem Z_2\rangle$ and
$\>r_{M_2}(E(M_2)\sem Z_2)\leq t$.
Then the matroid $M'$ represented by $\big(E(M_1)\sem Z_1\big)\cup E(M_2)$
has path-width at most~$t$.
\end{lemma}
\begin{proof}
We form a path-decomposition $Y$ of~$M'$ by appending
$Y_1\sem(Z_1\cup E(M_2))$ after~$Y_2$.
Let the considered subspaces (of the projective space over~$\FF$) be 
$\Pi:=\langle E(M_1)\rangle\cap\langle E(M_2)\rangle$ and
$\Sigma:=\langle E(M_2)\sem Z_2\rangle\supseteq\Pi$.
Let $(X,X')$ be a bipartition of $M'$ displayed by~$Y$.
If $X\subseteq Z_2\subseteq E(M_2)$, then
$\langle X\rangle\cap \langle E(M_1)\rangle\subseteq\Pi$,
and since $\langle E(M')\sem X\rangle\supseteq\Sigma\supseteq\Pi$,
we have $\langle X\rangle\cap \langle E(M')\sem X\rangle=
 \langle X\rangle\cap \langle E(M_2)\sem X\rangle$ 
which is of rank $\leq t$ by the assumption $w_{M_2}(Y_2)\leq t$.

If $Z_2\subseteq X\subseteq Y_2$, then
$\langle X\rangle\cap \langle E(M')\sem X\rangle\subseteq
 \langle \Sigma\cup\Pi\rangle=\Sigma$
is easily of rank~$\leq t$.
In the remaining case of $Y_2\subseteq X$ we get,
similarly as in the first case,
$\langle X\rangle\cap \langle E(M')\sem X\rangle\subseteq
 \langle \Pi\cup(X\sem E(M_2))\rangle\cap \langle E(M')\sem X\rangle\subseteq
 \langle Z_1\cup(X\cap E(M_1))\rangle\cap \langle E(M_1)\sem(X\cup Z_1)\rangle$
which is of rank $\leq t$ by the assumption $w_{M_1}(Y_1)\leq t$.
\end{proof}

\section{Self-reduction Algorithm}\label{sec:selfred}

In this section we give our core result---%
a self-reduction routine that, for a fixed parameter~$t$, 
constructs an optimal path-decomposition of a given represented
matroid of path-width~$t$, using an oracle which can decide whether 
the path-width of a given matroid is at most~$t$.
We stress that our routine can work with {\em any} oracle (subroutine)
for deciding the path-width value,
and that it is not restricted to only representable matroids as we will see
in the next Section~\ref{sec:improve}.

\paragraph{Motivation}
For easier understanding of the problem we are dealing with, 
we start this section with a brief overview of
the algorithm for constructing an optimal branch-decomposition of a given
(represented) matroid of branch-width~$t$ from~\cite{ho08};
it is based on the following decision step:
\begin{itemize}
\item$\!\!$\cite{ho08}
Assume $X\subseteq E=E(M)$ is such that $\lambda_M(X)\leq t$
and that $M[X]$ (the restriction of $M$ to~$X$) has branch-width $\leq t$.
The task is to decide whether $M$ has a branch-decomposition of width
$t$ such that ``$X$ forms one branch'' of the decomposition.
\end{itemize}
The way this decision task is implemented in \cite{ho08} is based on
extending $M\sem X$ with a bounded number of elements so that
every optimal branch-decomposition of it displays a separation whose guts is
geometrically identical with that of $(X,E\sem X)$ of~$M$
(then a branch formed by $X$ can be simply added to this place).
Besides implementing this key decision task, the rest of the algorithm
of~\cite{ho08} is an easy recursive composition routine (merging branches
until the whole tree is constructed).
On a very high level, our new algorithm will do the same thing tailored to
path-width -- see next.
Though, the underlying details will be very different and more complicated
due to the fact that one {\em cannot} ``add a branch'' to a path-decomposition
as to a branch-decomposition.

\paragraph{Algorithm outline}
We give a high-level description of our new path-decom\-po\-sition algorithm.
We now treat a given matroid $M$ represented over a finite
field $\FF$ as a point configuration in a projective geometry over~$\FF$
(recall Section~\ref{sec:prelim}):
Let $M$ be the input matroid and $E=E(M)$, $n=|E|$, where the points of $E$ are
given as vectors over~$\FF$.
For a simplification of the arguments, we assume that $|\FF|\geq3$,
that is, if $M$ is given with a representation over $GF(2)$ then we
equivalently view it over~$\FF=GF(4)$.

Assume that $\pw(M)=t$.
\begin{enumerate}[(I)]
\item \label{it:testpath-start}
For $i=1,2,\dots,n$, suppose that we have got a sequence
$X=(e_1,\dots,e_{i-1})\in E^{i-1}$ such that there exists a path-decomposition
of $M$ of width $t$ which starts with the prefix~$X$
(note that initially $X=\emptyset$ and our assumption is trivial).
\item \label{it:testpath-f}
For each $f\in E\sem X$, we set $X_f=(e_1,\dots,e_{i-1},f)$.
If $\lambda_M(X_f)\leq t$, we test whether there exists a path-decomposition
of $M$ of width $t$ which starts with the prefix~$X_f$.
\item If the test of \eqref{it:testpath-f} succeeds for some (any)~$f$\,%
---which has to happen for at least one value by the assumption---we 
let $X:=X_f$ and continue with \eqref{it:testpath-start}.
\end{enumerate}
Clearly, this scheme results in the construction of a path-decomposition 
$(e_1,\dots,e_{n})$ of~$M$ of width~$t$.
Hence it remains to explain implementation of crucial Step~\eqref{it:testpath-f}.

For convenience, we refer by $X_f$ also to the underlying set 
of the sequence $X_f$ from the above outline.
Unlike in the easier case of~\cite{ho08}, it is now not sufficient to test
$M\sem X_f$ for path-width $\leq t$ under the condition that the guts of
the bipartition $(X_f,E\sem X_f)$ is geometrically identical to the guts of
some bipartition displayed by the corresponding optimal path-decomposition.
We actually need that the corresponding optimal path-decomposition of $M\sem X_f$
can be ``prefixed'' with this guts without increasing the width
(which could be impossible if the displaying bipartition is somewhere in the
middle of the decomposition).
This goal we achieve by adding to $M\sem X_f$ a special set $D$ of points of rank
$t+1$ and path-width~$t$ (in fact, $D$ is represented over an extension field of~$\FF$).
Denoting by $M'$ the new matroid on $(E(M)\sem X_f)\cup D$,
it is then easy to see that the path-width of $M'$ is $\leq t$ if the answer
to \eqref{it:testpath-f} is {\sc Yes}
(see Lemma~\ref{lem:repl-pw} with $M_1=M,Z_1=X_f$ and $M_2\sim D$).
Proving the converse of this claim constitutes the core of the proof below.

The formal details are given below, in Algorithm~\ref{alg:main} and its proof.

\begin{alg}\label{alg:main}\rm
Let $\FF$ be a fixed finite field and $t\in\mathbb N$ a fixed parameter.
Let $\cal P$ be an oracle which, given any matroid $N$ represented
over~$\FF$, correctly decides whether~$\pw(N)\leq t$.
Let $M$ be an input connected $n$-element matroid of rank $r$,
given as an $r\times n$ matrix~$\mx A$ over~$\FF$, and
assume~$\pw(M)=t$.
\begin{enumerate}
\item \label{it:stepA1}
We pad $\mx A$ with $0$'s to make an $(r+t+1)\times n$ matrix
(informally, adding ``extra dimensions'' useful in the computation).
For simplicity, we will refer to the columns of the matrix as to the
elements of $M$, with understanding that all computations will be carried
out by means of linear algebra (i.e., in the matrix) in a natural way.
\item \label{it:stepX2}
Let initially $X:=\emptyset$.
For $i=1,2,\dots,n$, we repeat the following instructions:
\begin{enumerate}
\item 
We have got $X=(e_1,\dots,e_{i-1})\in E(M)^{i-1}$ where the elements of the
sequence are distinct, and we use the symbol $X$ to refer both to the
sequence and the underlying set of elements of~$M$.
\item \label{it:step2-f}
We choose $f\in E(M)\sem X$ such that $\lambda_M(X\cup\{f\})\leq t$, 
and set $X_f:=(e_1,\dots,e_{i-1},f)$.
\item \label{it:step2-gamma}
We compute the guts $\Gamma:=\langle X_f\rangle\cap\langle E(M)\sem X_f\rangle$
and choose a subspace $\Sigma\supseteq\Gamma$ of rank exactly $t$
and an element $d_0\not\in\Sigma$,
such that $\langle \Sigma\cup\{d_0\}\rangle\cap\langle E(M)\rangle=\Gamma$.
(Note that the rank of $\Gamma$, by \eqref{it:step2-gamma},
may be smaller than $t$, and we use some of the ``extra dimensions'' from
Step \eqref{it:stepA1} for placing $d_0$ and $\Sigma\supseteq\Gamma$ of rank exactly $t$.)
Let $P$ denote the set of all points of $\Sigma$ in the finite
projective geometry over~$\FF$.
Specially, for $t=1$, we form $P$ by two parallel points.
\item \label{it:step2-fext}
Let $N_0$ denote the matroid of rank $t+1$ induced by the points of
$P\cup\{d_0\}$, and $\FF_0=\FF$.
For $j=1,2,\dots,t$, let $N_j$ be the matroid constructed as a free extension of
$N_{j-1}$ by an element~$d_j$.
By Lemma~\ref{lem:extalpha}, $N_j$ is represented over the
extension field $\FF_j$ obtained from $\FF_{j-1}$ by adjoining a root of
degree~$r(N_0)=t+1$.
At the end, let $D_0:=\{d_0,d_1,\dots,d_t\}$, $D:=P\cup D_0$ and $\FF'=\FF_t$.
\item \label{it:step2-oracle}
For the matroid $M'$ induced on the point set $(E(M)\sem X_f)\cup D$
in the projective geometry over~$\FF'$,
we ask the oracle $\cal P$ whether $\pw(M')\leq t$.
\begin{itemize}
\item If the answer is {\sc No}, then we repeat the Steps
from~\eqref{it:step2-f} for another choice of~$f$.
\item If the answer is {\sc Yes}, then we update $X:=X_f$ and continue the
cycle in Step~\eqref{it:stepX2} with the next value of $i$ until~$i=n$.
\end{itemize}
\end{enumerate}
\item 
We output the path-decomposition $X=(e_1,\dots,e_{n})$ of~$M$
of width~$t$.
\end{enumerate}
Note that, in Step~\eqref{it:step2-oracle}, some element $e$ of $M$ may
be in the guts of $(X_f,E\sem X_f)$ and then $e$ is represented by the same
point as some element of $P$ in~$M'$.
It actually does not matter whether we consider these two elements as
identical or a parallel pair.
\end{alg}

\begin{theorem}\label{thm:main}
Let $\FF$, $t$ and $\cal P$ be as in Algorithm~\ref{alg:main}.
For any connected $n$-element input matroid $M$ represented by a matrix over $\FF$,
such that $\pw(M)=t$, Algorithm~\ref{alg:main} correctly outputs
a path-decomposition of $M$ of width~$t$.
With fixed parameters $\FF$ and $t$, the algorithm computes in FPT time
$\OO(n^4)$ and, in addition, makes $\OO(n^2)$ calls to the oracle~$\cal P$.
\end{theorem}

\begin{proof}
We start with justifying correctness of the algorithm.
Thanks to the condition $\lambda_M(X\cup\{f\})\leq t$ 
in Step \eqref{it:step2-f} of Algorithm~\ref{alg:main}, 
we know that the (eventual) output of 
the algorithm must be a path-decomposition of $M$ of width~$t$.
Consequently, it is enough to prove that for every iteration of Step
\eqref{it:stepX2} there is a choice of $f\in E(M)\sem X$ which correctly
succeeds in the test of Step \eqref{it:step2-oracle}.
Assuming, for this moment, the following
\begin{claim}\label{claim:M'}\rm
in Step~\eqref{it:step2-oracle}, $\pw(M')\leq t$ if and only if there exists a path-decomposition
of $M$ of width $t$ which starts with the prefix~$X_f$, 
\end{claim}\noindent
the rest of the proof follows by a straightforward induction on~$i$.

\medskip
It is hence enough to prove Claim~\ref{claim:M'}.
In one direction ($\Leftarrow$), assume that there exists a path-decomposition
$Y=(e_1,\dots,e_{n})$
of $M$ of width $t$ which starts with the prefix~$X_f$.
We give a path-decomposition $Y'=(e'_1,e'_2,\dots)$
of the matroid $N_t$ induced by the point set~$D$,
where $e'_1=d_0, e'_2=d_1, \dots, e'_{t+1}=d_t$ and this is followed 
by the elements of $P$ in any order.
The bipartition at position $j+1$ in $Y'$, for $j<t$, has the guts
$\langle\{d_0,\dots,d_j\}\rangle$ of rank $\leq t$.
At positions $j+1$ for $j\geq t$, on the other hand,
the guts is always $\Sigma$ of rank~$t$ (or its subspace).
Therefore, we can set $M_1=M, Z_1=X_f$ and $M_2=N_t, Z_2=D_0$
and apply Lemma~\ref{lem:repl-pw}, to conclude that $\pw(M')\leq t$.

\medskip
In the opposite direction ($\Rightarrow$) of Claim~\ref{claim:M'}, 
we assume that $\pw(M')\leq t$.
Recall the set $P$ of the points of $\Sigma$ over~$\FF$ from
Step~\eqref{it:step2-gamma}, and
the matroid $M'$ on the point set $(E(M)\sem X_f)\cup P\cup D_0$
from Step~\eqref{it:step2-oracle}.
Let $Y'=(e'_1,\dots,e'_p)$ be an optimal path-decomposition of~$M'$
where $p=|E(M')|$.
We first aim to show that there exists an index $1\leq j\leq p$ such that
the guts at the position  
$j$ in $Y'$ contains~$\Sigma$ (and so it equals~$\Sigma$ and $\pw(M')=t$).
If $t=1$, then $P$ consists of two parallel points (parallel to
single-point~$\Sigma$) and we simply choose a position between those points.
For $t>1$ this conclusion follows from Lemma~\ref{lem:alwaysSigma}
applied to the restriction of $Y'$ onto~$P$.

Let $Y'_j=\{e'_1,\dots,e'_j\}$ where~$\lambda_{M'}(Y'_j)=t$ by the previous
paragraph.
Recall the point set $D_0=\{d_0,\dots,d_t\}$ from Step~\eqref{it:step2-fext}.
We first claim that, up to possible reversal of the sequence $Y'$,
we have $D_0\subseteq Y'_j$.
This easily follows from the conclusion of Lemma~\ref{lem:onesidesep}
since, for any $0\leq a<b\leq t$, we have
$\big\langle\Sigma\cup\{d_a\}\big\rangle=\big\langle\Sigma\cup\{d_b\}\big\rangle
 =\big\langle\Sigma\cup D_0\big\rangle$ of rank $t+1$,
and so the condition of the lemma
$\mu_{M'}(Y'_j,\{d_a\})= \mu_{M'}(Y'_j,\{d_b\})=
 \mu_{M'}(Y'_j,\{d_a,d_b\})= t+1$ holds true.

Second, we claim that
 $(E(M)\sem X_f)\cap Y'_j\subseteq \langle P\rangle=\Sigma$. 
Suppose the contrary, that 
$Z:=\big((E(M)\sem X_f)\cap Y'_j\big)\sem \Sigma\not=\emptyset$
(here we view $E(M)\sem X_f$ as points in a projective space).
Note that $Z\cap\Sigma=\emptyset$.
We first consider the subcase that $\langle Z\rangle\cap\Sigma\not=\emptyset$.
Informally, we are going to argue that the spans of $Z$ and $D_0$ 
``freely overlap'' in $\Sigma$ and so, 
for any $g\in Z\cup D_0$, the span of $(Z\cup D_0)\sem\{g\}$ still contains
$\langle D_0\rangle$.
Then the path-decomposition $Y'$, at some position before $j$, must
contain $\langle D_0\rangle$ in the guts, but this is impossible since the
rank of $D_0$ is~$t+1$.
The corresponding formal argument follows.

Let $M''=M'\sem D_0$.
We choose $Z_0\subseteq Z$ minimal by inclusion such that
$\langle Z_0\rangle\cap\Sigma\not=\emptyset$,
and so the rank of $\langle Z_0\rangle\cap\Sigma$ is one
(in matroid terms this reads $r_{M''}(Z_0)+r_{M''}(P)=r_{M''}(Z_0\cup P)+1$\,).
Since $P$ contains all the points of $\Sigma$ in the projective geometry
over~$\FF$ (in matroid terms, $P$ is a modular flat in~$M''$ which is
represented over $\FF$),
we have that $\langle Z_0\rangle\cap P\not=\emptyset$,
and by minimality of $Z_0$ we have $\langle Z_0\rangle\cap P=\{p_0\}$.
Consequently, $C_0=Z_0\cup\{p_0\}$ is a circuit in $M''$ and so also in~$M'$.
Now we look at the set $C_1:=D_0\cup\{p_0\}$ in $M'$
which is of rank $t+1$ and cardinality $t+2$, and hence is dependent.
Since $d_1,\dots,d_t$ have been chosen as free extensions in Step
\eqref{it:step2-fext}, there cannot be any smaller circuits in $C_1$ and so
$C_1$ itself is a circuit.
We apply Lemma~\ref{lem:circexch-exact} to $C_0$ and $C_1$,
obtaining a circuit $C_2:=C_0\Delta C_1$ of $M'$, where no element of $C_2$
belongs to $\Sigma$ (our guts at position~$j$).
Since $C_2\supseteq D_0$, the span of $C_2$ contains $\Sigma$
and so $\mu_{M'}(Y'_j,C_2)=r_{M'}(C_2)$ and the conditions of
Lemma~\ref{lem:circuitspan} are fulfilled for~$C_2$.
However, the conclusion of the lemma contradicts our assumption $C_2\subseteq Y'_j$.

Next, still under the assumption $Z\not=\emptyset$, we consider the subcase that
$\langle Z\rangle\cap\Sigma=\emptyset$.
Recall that $\Gamma=\langle X_f\rangle\cap\langle E(M)\sem X_f\rangle=
 \langle \Sigma\cup D_0\rangle\cap\langle E(M)\rangle$.
Let $Z':=(E(M)\sem X_f)\sem Y'_j$ and note that $E(M)\sem X_f\supseteq
 Z\cup Z'\supseteq (E(M)\sem X_f)\sem\Sigma=(E(M)\sem X_f)\sem\Gamma$.
Hence, $\langle X_f\rangle\cap\langle \Gamma\cup Z\cup Z'\rangle=\Gamma$
and so $\langle Z\rangle\cap\langle X_f\cup \Gamma\cup Z'\rangle=
 \langle Z\rangle\cap\langle \Gamma\cup Z'\rangle$.
Since $M$ is connected, we in particular have
$\emptyset\not=\langle Z\rangle\cap\langle \Gamma\cup X_f\cup Z'\rangle=
 \langle Z\rangle\cap\langle \Gamma\cup Z'\rangle$.
The latter in turn means, again from the path-decomposition $Y'$ of~$M'$
with the guts $\Sigma\supseteq\Gamma$ at position~$j$, that
$\langle Z\rangle\cap\Sigma\not=\emptyset$ -- the case already being considered above.

To recapitulate, the assumed path-decomposition $Y'$ of $M'$ has
the (geometric) guts $\Sigma=\langle Y'_j\rangle\cap\langle E(M')\sem
 Y'_j\rangle$ at position~$j$. 
We have also shown that $Y'_j\cap(E(M)\sem X_f) \subseteq \Sigma$.
Hence, if we form $Y''$ by restricting $Y'$ to the elements of
$E(M)\sem X_f$, the concatenated sequence $(X_f,Y'')$ will be
a path-decomposition of $M$ of width~$t$.
The proof of Claim~\ref{claim:M'} is finished.

\medskip
The last point is to address runtime complexity of Algorithm~\ref{alg:main}.
Note that the finite field $\FF$ and the value of $t$ are fixed parameters.
In particular, arithmetic operations over $\FF$ and $\FF'$
(which depends only on $\FF$ and $t$) take constant time each.
Also note that $r\leq n$.
We $n$ times iterate at Step \eqref{it:stepX2}, and each iteration costs the following.
We are choosing at most $n$ values of $f$ in Step~\eqref{it:step2-f}, and for
each we compute the subspace $\Gamma$.
Knowing $\langle X\rangle\cap\langle E(M)\sem X\rangle$ already from the
previous level, the computation of
$\Gamma=\langle X_f\rangle\cap\langle E(M)\sem X_f\rangle$
takes $\OO(n^2)$ in Step~\ref{it:step2-fext}
by standard linear algebra (the rank of $\Gamma$ is at most constant~$t$).
The rank of $\Sigma\supseteq\Gamma$ and cardinality of the set $P$ 
are constants depending on $\FF$ and $t$.
Step~\eqref{it:step2-fext} takes $\OO(1)$ time since it depends only on $\FF$
and $t$ and not on the input~$M$.
In fact, the point set $D_0$ needs to be computed only once during the
whole algorithm and then linearly transformed to match actual~$\Sigma$.
This amounts to $\OO(n^4)$ total time and $\OO(n^2)$ calls to the oracle
$\cal P$ in Step~\eqref{it:step2-oracle}.
\end{proof}

\section{Algorithmic Consequences}\label{sec:improve}

So far, in Section~\ref{sec:selfred} we have restricted attention to
connected matroids, but this is not any problem since we may easily
concatenate {path-decompositions of connected components} of a general
matroid.
To make use of Algorithm~\ref{alg:main}, we also need to provide an
implementation of the oracle~$\cal P$
(which tests the value of path-width~$\leq t$, as sketched in the Introduction).
This will be done by Theorem~\ref{thm:n3test}.
A class $\scri N$ of matroids is {\em minor-closed} if, for every matroid
$M\in\scri N$, all minors of $M$ also belong to~$\scri N$.
A matroid $M\not\in\scri N$ is an {\em obstruction for membership in}
$\scri N$ if all proper minors of $M$ belong to~$\scri N$.

\begin{theorem}[Geelen--Gerards--Whittle~\cite{ggw02},%
\footnote{%
We remark that Geelen, Gerards and Whittle have announced a ``matroid minors''
theorem which does not require a bound on branch-width to claim finite
number of $\FF$-representable obstructions for $\scri N$, but that is not
fully published yet. For our purpose, the version of~\cite{ggw02} is sufficient.
}
		and Hlin\v{e}n\'y~\cite{hli06}]
\label{thm:n3test}
Let $\FF$ be a fixed finite field and $k\in\mathbb N$ a fixed parameter.
For any minor-closed class $\scri N$ of matroids,
there are finitely many obstructions for membership in $\scri N$
which are representable over $\FF$ and have branch-width at most~$k$.
Consequently, there is an FPT algorithm which,
given an $n$-element matroid $M$ represented by a matrix over $\FF$, 
in time $\OO(n^3)$ correctly decides whether~$M\in\scri N$ 
or outputs that the branch-width of $M$ is more than~$k$.
\end{theorem}

\paragraph{Direct implementation}
The way we use Theorem~\ref{thm:n3test} in an implementation of the oracle
$\cal P$ combines Lemma~\ref{lem:pwbw} with Lemma~\ref{lem:pwminor};
the matroids of path-width at most $t$ have branch-width at most~$t+1$
and form a minor-closed class 
$\scri P_t$ for which we can test membership in FPT time~$O(n^3)$.
Note, though, that this approach results in a nonuniform FPT algorithm
since we do not explicitly know the finite lists of obstructions for the
classes $\scri P_t$,~$t\in\mathbb N$.
In combination with Theorem~\ref{thm:main} we immediately get:

\begin{corollary}\label{cor:direct-n5}
Let $\FF$ be a fixed finite field and $t\in\mathbb N$ a fixed parameter.
There is a nonuniform FPT algorithm parameterized by $t$ and $|\FF|$ which,
given an $n$-element matroid $M$ represented by a matrix over $\FF$,
in time $\OO(n^3)$ decides whether~$\pw(M)\leq t$.
\\
Consequently, if $\pw(M)\leq t$, 
there is a nonuniform FPT algorithm parameterized by $t$ and $|\FF|$, 
which in time $\OO(n^5)$ outputs
a path-decomposition of $M$ of width~$t$.
\qed\end{corollary}

We remark that, in the setting of nonuniform algorithms,
Algorithm~\ref{alg:main} as used in Corollary~\ref{cor:direct-n5} can be
further simplified by the following observation.
The point configuration $D$ constructed in Step~\eqref{it:step2-fext} is
unique, up to a linear transformation, for given parameters $t,\FF$,
and hence it can be hard-coded into the (anyway nonuniform) algorithm
with the smallest possible extension field~$\FF'$ which can represent~$D$
(this would quite likely be a much smaller field than the one computed by
brute force in Step~\eqref{it:step2-fext}).

\paragraph{Improving runtime}
Runtime dependence on $n$ of the algorithm of Corollary~\ref{cor:direct-n5} can be
improved to $\OO(n^3)$ by using the same implementation tricks as
in~\cite{ho08}, based on earlier~\cite{hlineny05}.

\begin{theorem}\label{thm:alg-n3}
Let $\FF$ be a fixed finite field and $t\in\mathbb N$ a fixed parameter.
There is a nonuniform FPT algorithm parameterized by $t$ and $|\FF|$ which,
given an $n$-element matroid $M$ represented by a matrix over $\FF$,
in time $\OO(n^3)$ outputs
a path-decomposition of $M$ of width~$t$ or certifies that~$\pw(M)>t$.
\end{theorem}

\begin{proof}[Proof sketch]
In the improved algorithm, we follow the general scheme of 
\cite[Section~6]{ho08} but in a simplified way.
This is possible thanks to the fact that Algorithm~\ref{alg:main}, at each
iteration, works with only one ``active guts'' of a bipartition $(X,E\sem X)$,
unlike the algorithm of \cite{ho08} which builds many branches of the desired
branch-decomposition concurrently.

We modify the main steps of Algorithm~\ref{alg:main} as follows:
\begin{enumerate}
\item 
For the input matroid $M$ represented by the matrix $\mx A$,
we use \cite{hlineny05} to compute a branch-decomposition of $M$ of width
at most $3t+3$\,---actually, a so-called {\em$3t$-boundaried parse tree} $\ca T$ for
$M$\,---or to confirm that $\bw(M)>t+1$ and so $\pw(M)\geq\bw(M)-1>t$.
This step takes $\OO(n^3)$ time for fixed~$t,\FF$.
\item
Each task performed in Steps \ref{it:step2-gamma} and \ref{it:step2-fext}
can be done in time $\OO(n)$ within the parse tree $\ca T$
(we refer to \cite[Section~6]{ho08} for corresponding details).
It is important to compute within $\ca T$ (and not on whole $\mx A$), for
which purpose we each time ``enlarge'' every node of $\ca T$ by the 
constant-rank subspace~$\Sigma$.
Subsequently, Step \ref{it:step2-oracle} can test $\pw(M')\leq t$
by checking (non-)presence of the finitely many obstructions for `path-width
$\leq t$\,'.
This test can also be done in time $\OO(n)$ by \cite{hli06}
since minor obstructions are MSO-definable.
\end{enumerate}
Altogether, runtime is $\OO(n^3+n^2\cdot n)=\OO(n^3)$ for fixed~$t,\FF$.
\end{proof}

\paragraph{Abstract matroids}
Besides its simplicity, our Algorithm~\ref{alg:main} has another
theoretical advantage over the constructive algorithm of~\cite{jko16j}.
While the authors of \cite{jko16j} directly compute with points and subspaces in a finite
projective geometry, and it does not seem possible to extend their approach
to infinite projective geometries or abstract matroids, 
we can easily adapt our algorithm to
work even with {\em abstract matroids} given by a rank oracle
(although our algorithm also directly worked with the points of a subspace
$\Sigma$, that was only for convenience and clarity, and could be rather easily 
replaced by an abstract handling).

In this respect we mention the algorithm of Nagamochi~\cite{nag12}
which computes an optimal path-decomposition for an arbitrary submodular
function (and hence including the case of a matroid given by a rank oracle).
Though, its runtime is of order $\OO(n^{f(t)})$ where $t$ is the path-width
(complexity class XP) while we aim for an FPT algorithm.

We say that an abstract matroid $M$ is given by a {\em rank oracle}
if the input consists of the ground set $E=E(M)$ and an oracle function
${\cal R}: 2^E\to\mathbb N$ such that ${\cal R}(X)=r_M(X)$ for all~$X\subseteq E$.
Algorithms then handle $M$ by asking $\cal R$ so called {\em rank queries}.
In this setting we have got the following algorithm.

\begin{theorem}\label{thm:alg-abstract}
Let $t\in\mathbb N$
and $\cal P$ be an oracle function which, for any matroid $N$ given by a rank oracle,
correctly decides whether~$\pw(N)\leq t$.
There is an algorithm that,
for an input $n$-element matroid $M$ given by a rank oracle $\cal R$,
outputs a path-decomposition of $M$ of width~$t$ or correctly answers that~$\pw(M)>t$.
The algorithm makes $\OO(n^2)$ calls to the oracle function~$\cal P$
and, neglecting the fixed parameter $t$, asks $\OO(n^2)$ rank queries.
\end{theorem}

Before moving onto the proof, we need one more technical concept.
We are going to modify the matroid $M$ (which we do not completely know---we 
cannot read all the ranks of sets in $M$ in polynomial time!).
Instead, we will modify the rank oracle $\cal R$ by prescribing its
(efficient) answers to rank queries involving elements which we add to $M$.
In this respect we define the following three elementary operations:
\begin{enumerate}[(M1)]
\item \label{it:acoloop}
{\em Adding a coloop} $a$ to $M$ defines, for every $X\subseteq E(M)$,
that $\ca R(X\cup\{a\}):=\ca R(X)+1$.
\item \label{it:aclosure}
{\em Placing $b$ freely into the closure of $Z\subseteq E(M)$} defines, 
for every $X\subseteq E(M)$,
\begin{itemize}
\item $\ca R(X\cup\{b\}):=\ca R(X)$ if $r_M(Z\cup X)=r_M(X)$, and
\item $\ca R(X\cup\{b\}):=\ca R(X)+1$ otherwise.
\end{itemize}
\item \label{it:aguts}
{\em Placing $c$ freely into the guts of} (the bipartition of) 
$Z\subseteq E(M)$ means, for $X\subseteq E(M)$,
\begin{itemize}
\item $\ca R(X\cup\{c\}):=\ca R(X)$ if $\mu_M(Z,X)=r_M(X)$, and
\item $\ca R(X\cup\{c\}):=\ca R(X)+1$ otherwise.
\end{itemize}
\end{enumerate}

An informal geometric explanation of these operations follows.
(M\ref{it:acoloop}) simply ``adds another dimension'' with~$a$.
(M\ref{it:aclosure}) puts the new point $b$ in general position (i.e.,
without unforced linear dependencies) into the span $\langle Z\rangle$.
(M\ref{it:aguts}) similarly puts the new point $c$ in general position
into the guts $\langle Z\rangle\cap\langle E(M)\sem Z\rangle$.
It is a routine exercise to prove that the rank oracle defined
by each one of (M\ref{it:acoloop}), (M\ref{it:aclosure}), (M\ref{it:aguts})
is the rank function of a matroid.

\vspace*{-1ex}
\begin{proof}[Proof of Theorem~\ref{thm:alg-abstract}]
Again, we may restrict our attention to connected input matroids~$M$.
We modify some steps of Algorithm~\ref{alg:main} as follows:
\begin{itemize}
\item
Step \eqref{it:stepA1} is not needed.
\item
In step \eqref{it:step2-gamma}, let $k=\lambda_M(X_f)$.
First, we $(t-k)$-times (if $k<t$) repeat the operation (M\ref{it:acoloop}) 
of adding a coloop.
Let $P_0$ denote the set of coloops added to $M$ this way.
We then $(t+k)$-times repeat the operation (M\ref{it:aguts})
of placing a new element freely into the
guts of $\big(X_f\cup P_0,\, (E(M)\sem X_f)\cup P_0\big)$\,---to 
be formally precise, we consider for this operation the elements of $P_0$ duplicated.
Let $P\supseteq P_0$ denote the set of all the $2t$ added elements, which is
of rank~$t$ (one may observe that $P$ actually induces a uniform matroid $U_{t,2t}$).
\item
In Step \eqref{it:step2-fext}, we add a new coloop $d_0$ by (M\ref{it:acoloop}).
Then, for $j=1,\dots,t$, we iteratively do the operation (M\ref{it:aclosure})
of freely placing a new element $d_j$ into the closure of $P\cup\{d_0\}$.
Again, let $D_0:=\{d_0,d_1,\dots,d_t\}$ and $D:=P\cup D_0$.
\item
In Step \eqref{it:step2-oracle}, we let $M'$ be the matroid
defined on the ground set $(E(M)\sem X_f)\cup D$ by the rank oracle $\ca R'$
constructed from $\ca R$ by the above modifications.
\end{itemize}

\smallskip
In the proof of the modified algorithm, we can essentially repeat 
the setup and most of the arguments of the proof of Theorem~\ref{thm:main}, 
translated into the abstract setting of the rank functions of $M$ and $M'$.
Such as, geometric span of points representing $M'$ is translated as
the closure operation in $M'$ and, in particular,
the subspace $\Sigma=\langle P\rangle$ is now written as $cl_{M'}(P)$.
Though, the following two steps in the proof of the forward direction of
Claim~\ref{claim:M'} need separate formal arguments:
\begin{itemize}
\item 
Assuming a path-decomposition $Y'=(e'_1,\dots,e'_p)$ of~$M'$ of
width~$t$, we, instead of invoking Lemma~\ref{lem:alwaysSigma}, 
argue simply as follows:
We define index $j$ as the minimum $1\leq j\leq t$ such that $|P\cap Y'_j|=t$.
Since the elements of $P$ have been each freely placed into a rank-$t$ flat, 
the $t$-element set $P\cap Y'_j$ is independent, and so is the complement $P\sem Y'_j$.
Consequently, all elements of $P$ belong to the guts of the bipartition
at the position $j$ of~$Y'$, a situation analogous to the former proof.
\item
Second, we differently argue that
$(E(M)\sem X_f)\cap Y'_j\subseteq cl_{M'}(P)$.
Assuming $Z:=\big((E(M)\sem X_f)\cap Y'_j\big)\sem cl_{M'}(P)\not=\emptyset$,
we again aim to find a circuit $C_2\subseteq Y'_j$ 
contradicting the conclusion of Lemma~\ref{lem:circuitspan}.
A full proof of the existence of such $C_2$ is left for coming
Lemma~\ref{lem:getcircuitD} (in which $X=X_f$ and $C_2=D_0\cup Z_0$).
\end{itemize}
Assuming now Lemma~\ref{lem:getcircuitD}, the proof is finished.
\end{proof}

\begin{lemma}\label{lem:getcircuitD}
Let $M$ be a connected matroid, $X\subseteq E=E(M)$ and 
$\emptyset\not=Z\subseteq E\sem X$.
Assume that $P\subseteq E\sem X$ is such that $r_M(P)=t\geq\lambda_M(X)$,
$\mu_M(X,P)=t$, and $Z\cap cl_M(P)=\emptyset$.
Furthermore, assume that $M_0$ is a matroid on the ground set $E\cup D_0$
where $D_0=\{d_0,d_1,\dots,d_t\}$, the restriction of $M_0$ to $E$ is~$M$,
and $d_0$ is a coloop w.r.t.~$M$ and each $d_i$ is freely placed
(M\ref{it:aclosure}) in the closure
of $P\cup\{d_0\}$ w.r.t. $E\cup\{d_0,\dots,d_{i-1}\}$ for $i=1,\dots,t$.
Let $M'$ be $M_0$ restricted to $E(M_0)\sem X$.
If $\lambda_{M'}(D_0\cup Z)\leq t$, then there exists $Z_0\subseteq Z$ 
such that $D_0\cup Z_0$ is a circuit of~$M'$.
\end{lemma}

\begin{proof}
Let $E'=E\sem Z$.
Observe that $|D_0|=r_{M'}(D_0)=t+1$
 (since $D_0$ is inde\-pendent both in $M_0$ and $M'$),
$\lambda_{M'}(D_0)=t$ and $cl_{M'}(D_0)\supseteq P$, but
$D_0\cap cl_{M'}(P)=\emptyset=D_0\cap cl_{M'}(E')$.
From the assumptions $\lambda_{M'}(D_0\cup Z)\leq t$ and $Z\cap P=\emptyset$
we get that actually $\lambda_{M'}(D_0\cup Z)=t=r_{M'}(P)$ 
and all elements of $P$ are in the guts of
$(D_0\cup Z,E'\sem Z)$ in~$M'$.

Let $e_1\in X$ and $e_2\in Z$ be arbitrary.
Since $M$ is connected, there exists a circuit $C\subseteq M$, $C\ni e_1,e_2$.
We apply Lemma~\ref{lem:circguts} to $M$, $C$ and $Y:=X$, $Q:=P$, $e:=e_2$.
The obtained circuit $C'$ satisfies:
$e_2\in C'\subseteq (C\sem X)\cup P\subseteq E(M')$ 
and $C'\subseteq cl_{M'}(E')$.

In the matroid~$M'$, we let $Y:=cl_{M'}(E'\sem Z)\supseteq P$.
From previous $\lambda_{M'}(D_0\cup Z)=r_{M'}(P)$ where $P$ is in the guts,
we have $Z\cap Y\subseteq cl_{M'}(P)$;
and since $Z\cap cl_{M'}(P)=\emptyset$ by the assumptions,
we then get $Z\cap Y=\emptyset$.
In the matroid $M''=M'/d_0$ obtained by contracting $d_0$,
we have $D_1=\{d_1,\dots,d_t\}\subseteq cl_{M''}(P)$
and $cl_{M''}(D_1)=cl_{M''}(P)$,
since $cl_{M'}(D_0)\supseteq P$ and $r_{M''}(D_1)=t=r_{M''}(P)$.
Note that $C'$ is a circuit of $M''$, too,
since $M''$ restricted to $E'$ equals $M\sem X$.
Denoting $Y'=E(M'')\sem Y$, we have $e_2\in Z\cap C'\subseteq Y'\cap C'$.
In this setup, we apply Lemma~\ref{lem:circguts} to $M''$, $Y$ and
$C:=C'$, $Q:=D_1$, $e:=e_2$.
The circuit $C''$ that we obtain, satisfies
$C''\subseteq (C'\sem Y)\cup D_1$, and so $C''\subseteq Z_0\cup D_1$ 
where $Z_0=Z\cap C''$ (since $Y\supseteq E'\sem Z$).

Back in the matroid $M'$ (uncontracting $d_0$), $C''\cup\{d_0\}$ is a circuit 
of $M'$, and $C''\supseteq D_1$ since the elements of $D_1$ have been freely
placed---they do not have unforced dependencies in $M'$.
Hence this circuit is $D_0\cup Z_0=C''\cup\{d_0\}$.
\end{proof}

\section{Conclusions}\label{sec:conclu}

We have shown a relatively simple oracle algorithm which can construct an
optimal path-decomposition of a given matroid if it is provided with a
subroutine testing the value of matroid path-width.
This completes the picture of width decompositions of ($\FF$-represented) matroids
in the following sense:
While for the matroid branch-width, a non-constructive FPT decision algorithm
has been known since \cite{hlineny05}, followed by a natural self-reduction
constructive algorithm in~\cite{ho08}, no such FPT self-reduction approach to
constructing an optimal matroid path-decomposition seemed possible 
along similar lines before.

Specifically for matroids represented over a finite field $\FF$, this result 
provides an alternative to the 
recent algorithm of Jeong, Kim and Oum~\cite{jko16,jko16j}
which uses a direct and complicated construction based on ideas originally 
developed for graphs by Bodlaender and Kloks~\cite{bk96}.
Though, there is price we have to pay for simplicity of our algorithm;
our approach provides a nonuniform FPT algorithm, caused by the fact that we
have yet no explicit bound on the size of the minor-minimal obstructions for
path-width~$\leq t$ (unlike the case of branch-width in which
an explicit bound~\cite{ggrw03} readily provides a uniform FPT
algorithm~\cite{ho08}).

Moreover, our self-reduction oracle algorithm readily generalizes
to abstract matroids given by rank oracles, as proved in Theorem~\ref{thm:alg-abstract}.
Although we are currently not aware of an FPT algorithm which could test
path-width~$\leq t$ for matroids given by rank oracles,
such algorithms could probably emerge in the future (cf.~\cite{nag12}) 
for other matroid classes,
and then Theorem~\ref{thm:alg-abstract} will be readily applicable also to
these new classes.
Along the same line, it is likely that in the future an
explicit bound on the obstructions for path-width~$\leq t$
will be found and then Algorithm~\ref{alg:main} will immediately turn uniform.

\subsection*{Acknowledgments}
We would like to thank the anonymous referees for careful reading and
checking all the proofs, and for many suggestions which helped  to improve
this paper a lot.

\bibliographystyle{plain}\small
\bibliography{mpw-self-arxivf}

\end{document}